\documentclass[twocolumn,showpacs,prb]{revtex4}%
\usepackage{epsfig}
\usepackage{caption}
\usepackage{lscape}
\usepackage{bm}
\usepackage{amsfonts}
\usepackage{amsmath}
\usepackage{amssymb}
\usepackage{graphicx}%

\input{tcilatex}

\begin{document}

\title{Theory of resonant spin Hall effect}
\author{ Fu-Chun Zhang and Shun-Qing Shen}
\date{January 20, 2007}

\begin{abstract}
A biref review is presented on resonant spin Hall effect, where a tiny
external electric field induces a saturated spin Hall current in a
2-dimensional electron or hole gas in a perpendicular magnetic field. The
phenomenon is attributted to the energy level crossing associated with the
spin-orbit coupling and the Zeeman splitting. We summarize recent
theoretical development of the effect in various systems and discuss
possible experiments to observe the effect.
\end{abstract}

\affiliation{Department of Physics and Center of Theoretical and Computational Physics ,
The University of Hong Kong, Pukfulam Road, Hong Kong, China}
\maketitle

\section{Introduction}

Spin-orbit interaction of electrons as a relativistic quantum mechanical
effect plays important roles in a number of classes of metals and
semiconductors. Spin-orbit coupling mixes different spin states and provides
an efficient way to control the coherent motion of electron spins via an
applied electric field. Directional motion of electron spins may circulate a
spin current with a null charge current. Spin Hall effect refers to spin
current or spin accumulation in the direction transverse to the applied
electric field, in analogy to charge current in the Hall effect. The spin
Hall effect was first proposed by Dyakonov and Perel in early of 1970's \cite%
{Dyakonov71}. The interest of the effect was revived in recent years\cite%
{Hirsch99,Zhang00prl,Hu03prb}. The "intrinsic" spin Hall effect was proposed
in 2003 by Murukami et al.\cite{Murakami03Science} for hole doped
semiconductors and by Sinova et al. \cite{Sinova04} for two dimesnional
electron gas with a Rashba spin-orbit coupling. Since then studies of the
spin Hall effect has evolved into a subject of intense research. Although
there are debates on the precise definition of the spin current, the detect
of a pure spin current continues to be a challenging issue. The spin current
may lead to the formation of spin accumulation or non-uniform spin
distribution at the boundaries, which may be detected experimentally. \cite%
{Kato04Science,Wunderlich05prl,Sih05NP} The spin current may also lead to a
Hall voltage or charge Hall current as a reciprocal effect induced by spin
current, which has been reported in diffusive metallic conductors.\cite%
{Valenzuela06,Saitoh06apl} In addition, quantum interferences by optical
means have also been reported.\cite{Hubner03prl,Stevens03prl,Zhao06prl}
Recently it was observed that an optically injected spin current flowing
through a Hall-bar system can generate an in-ward or out-ward electric
current, while the Hall voltage remains zero.\cite{Cui06} This observation
renders a manifestation of the tensor-like nature of a spin current, of
which both the spin polarization and the velocity are decisive factors in
producing observable effects.

Resonant spin Hall effect refers to a resonant or saturated spin Hall
current rsponse to a tiny applied electric field in a 2DEG with spin-orbit
coupling in the presence of a strong perpendicular magnetic field. As a
result the spin Hall conductance may become divergent in a weak field limit.
The resonance was attributed to the energy level crossing of the system due
to the competition between the spin-orbit coupling and the Zeeman splitting
or other terms. The resonant spin Hall effect was first proposed by Shen et
al. in a 2DEG with a Rashba coupling\cite{Shen04prl,Shen05prb}. The edge
spin current was studied by Bao et al. \cite{Bao05prb}. The resonant spin
Hall effect in a hole-doped system described by a Luttinger hamiltonian has
recently been studied by Zarea and Ulloa \cite{Zarea06prb} and by Ma and Liu 
\cite{Ma06prb}. At persent, there have not been experimental reports on the
observation of the resonant spin Hall effect or related phenomena yet, which
is likely due to the combination of the difficulty in detecting the spin
current or spin accumulation in the high magnetic field and the lack of
experimental efforts in looking into these phenomena.

In this review we shall summarize recent theoretical works on the resonant
spin Hall effect and relevant works on 2-dimesnional electron or hole gases.
For overviews on the general spin Hall effect, we refer readers to other
recent review articles and references therein.\cite{Murakami05,
Sinova06ssc,Schliemann06ijmpb} The rest part of the paper is organized as
follows. In section II, we review the resonant spin Hall effect in various
2-dimensional systems including the Rashba and Dresselhaus couplings of the
electron system and the hole system of the Luttinger Hamiltonian. In section
III, we review the edge spin current. In section IV, we discuss a close
relation between the spin Hall cuurent and the spin polarization and also
discuss the effect of the disorder. We propose experiments to observe the
resonant spin Hall effect.

\section{Resonant spin Hall effect}

In this section we discuss the theoretical studies of resonant spin Hall
effect (RSHE) in various systems. We will first discuss the effect in
2-dimesnional electron gas (2DEG) with a linear Rashba spin-orbit coupling,
followed by a discussion of the effect in 2DEG with Dresselhaus spin-orbit
coupling by examining a symmetry transformation to the Rashba coupling. We
then discuss the effect in 2DEG with both Rashba and Dresselhaus couplings
and the effect in 2-dimensional hole gas.

\subsection{RSHE in 2DEG with Rashba coupling}

The resonant spin Hall effect was first proposed in 2DEG with a Rashba
spin-orbit coupling $\beta $. The Hamiltonian for a sinlge electron of
spin-1/2 in 2DEG of area $L_{x}\times L_{y}$ subject to a perpendicular
uniform magnetic field $\mathbf{B}=B\hat{z}=\vec{\triangledown}\times 
\mathbf{A}$ and an in-plane electric field $\mathbf{E}=E\hat{y}$ is given by%
\begin{eqnarray}
H_{R} &=&\frac{1}{2m}\left( \mathbf{p}+\frac{e}{c}\mathbf{A}\right)
^{2}+\beta \hat{z}\cdot \left[ (\mathbf{p}+\frac{e}{c}\mathbf{A})\times
\sigma \right]  \notag \\
&&-\frac{g_{s}}{2}\mu _{B}B\sigma _{z}+eEy  \label{HamiltonianR}
\end{eqnarray}%
where $m,-e,g_{s}$ are the electron's effective mass, charge and the Lande
g-factor, respectively. $\mu _{B}$ is the Bohr magneton, and $\sigma
_{\gamma }$ ($\gamma =x,y,z$) are the Pauli matrices. We choose a Landau
gauge $\mathbf{A}=yB\hat{x}$, and consider a periodic boundary condition in
the $x$ direction. We consider the case $E=0$ first. At $\beta =0$, the
solution of the above Hamiltonian is well known and the energy spectra are
given by evenly-spaced Landau levels $(n+1/2)\hbar \omega -g_{s}\mu
_{B}B\sigma _{z}/2$ for each spin state ($\omega =eB/mc$, n=0,1,2,...). At $%
\beta \neq 0$, the Rashba term hybridizes the two neighboring Landau levels
with opposite spins, and the problem can be solved analytically with the
energy levels given by\cite{Rashba60, Shen04prl}, 
\begin{equation}
\epsilon _{ns}=\hbar \omega \left( n+\frac{s}{2}\sqrt{(1-g)^{2}+8n\eta ^{2}}%
\right) \,,
\end{equation}%
where , $\eta =\beta ml_{b}/\hbar $, and $g=g_{s}m/2m_{e}$, with $m_{e}$ the
mass of a free electron and $l_{b}=\sqrt{\hbar c/eB}$ the magnetic length
with $s=\pm 1$, for $n\geq 1$; and $s=1$ for $n=0$. The eigenstate $%
\left\vert n,k,s\right\rangle $ has a degeneracy $N_{\phi }=L_{x}L_{y}eB/hc$%
, corresponding to $N_{\phi }$ quantum values of $p_{x}=k$. The
two-component wavefunction is given by

\begin{equation}
\left\vert n,k,s\right\rangle =\left( 
\begin{array}{c}
\cos {\theta _{ns}}\phi _{nk} \\ 
i\sin {\theta _{ns}}\phi _{n-1k}%
\end{array}%
\right)
\end{equation}%
where $\phi _{nk}$ is the eigenstate of the $n^{th}$ Landau level in the
absence of the Rashba interaction. Only two levels of $\phi _{nk}$ with spin
up and $\phi _{n-1k}$ with spin down are mixed together. For $n=0$, $\theta
=0$, and for $n \geq 1$, $\tan {\ \theta _{ns}}=u_{n}-s\sqrt{1+u_{n}^{2}}$,
with $u_{n}=(1-g)/\sqrt{8n}\eta $. One of the key features of the solution
is that the energy levels may cross, which introduces additional degeneracy
of the spectrum (see Fig. 1).

\begin{figure}[ptb]
\centerline{\epsfxsize=8cm\epsfbox{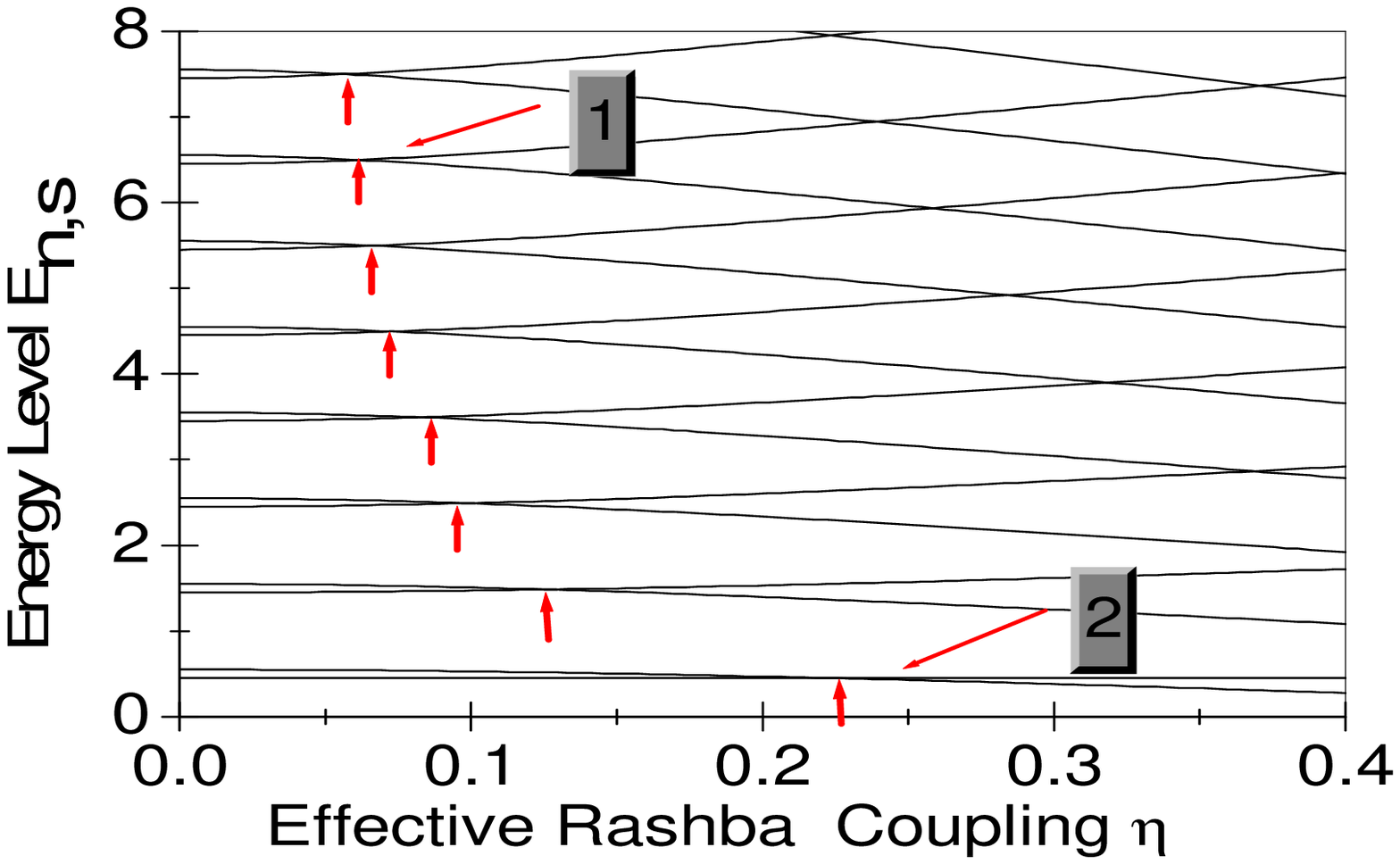}}\caption{
Landau levels of an electron as
functions of effective Rashba coupling $\protect\eta =\protect\alpha %
ml_{b}/\hbar $ for $g=g_{s}m/2m_{e}=0.1$. Arrows indicate those level
crossings giving rise to resonant spin Hall conductance in a weak electric
field. [From Ref.(\protect\cite{Shen04prl})]
}%
\label{figure.eps}%
\end{figure}

Analysis shows that the additional degeneracy caused by the crossing of the
Landau levels will lead to some anomalous properties related to electron
spins. One of them is the resonant spin Hall effect, where the spin Hall
conductance may become divergent when the Fermi surface is located at or
near the crossing point.\cite{Shen04prl} In other words, a tiny external
electric field $E$ will generate a saturated spin Hall current. This effect
can be best understood in a simplified two-level model\cite{Shen04prl}.

Let's consider two states $\left\vert 0,+1\right\rangle $ and $\left\vert
1,-1\right\rangle$ and the effective Hamiltonian $H_R$ in the presence of an
electric field $E$ is reduced to 
\begin{equation}
H_{\text{reduced}}=\left( 
\begin{array}{cc}
\Delta \epsilon & v_{0} \\ 
v_{0} & -\Delta \epsilon%
\end{array}%
\right)
\end{equation}%
where $\Delta \epsilon =(\epsilon _{0,+1}-\epsilon _{1,-1})/2,$ and $%
v_{0}=-eEl_{b}\eta\cos \theta _{0,+1}\sin \theta _{1,-1}.$ A key feature of
this model is that the electric field generates a non-zero off-diagonal
elements. Near the crossing point the electron spins in the two levels are
almost opposite. The external electric field hybridizes the two levels and
opens an energy gap. Each mixed state carries a finite spin Hall current
with the opposite direction. When the lower energy states are occupied, a
finite spin Hall current circulates once the energy gap is openned by the
external field. A detailed analysis shows that at zero temperature and weak
field limit, the spin Hall conductance $G_{s}=-\frac{e}{4\pi }\frac{\delta
_{\nu }B_{0}}{\left\vert B-B_{0}\right\vert }$ where $\delta _{\nu }$ is the
filling factor at the two levels. At low $T$, as the magnetic field
approaches the resonant point $B\rightarrow B_{0}$, $G_{s}^{z}\rightarrow
-e\nu \frac{\sqrt{2}}{4\pi }g^{2}(1+g)^{-2}\hbar \omega _{0}/kT$, and $\int
G_{s}^{z}dB\rightarrow -e\nu \sqrt{2}B_{0}g/4\pi (1+g)\ln \left[ \hbar
\omega _{0}/k_{B}T\right] $. At the resonant point and low temperature, $%
G_{s}\propto 1/E$.

\subsection{RSHE in 2DEG with Dresselhaus couping}

In III-V compounds such as GaAs and InAs, the spin-orbit interaction also
generates a Dresselhaus coupling in systems with bulk inversion asymmetry,
which is given by $H_{so}^{D}=\alpha (p_{x}\sigma _{y}-p_{y}\sigma _{x})$.
The Rashba couping and Dresselhaus coupling plays different roles in the
spin Hall effect. In this subsection, we examine the symmetries related to
the Rashba and Dresselhaus couplings. We consider a 2D spin-orbit coupling, $%
V_{so}^{2D}=H_{so}^{R}+H_{so}^{D}$.

\textit{Interchange symmetry of the two couplings. }Under the unitary
transformation, $\sigma _{x}\rightarrow \sigma _{y},$ $\sigma
_{y}\rightarrow \sigma _{x},$ $\sigma _{z}\rightarrow -\sigma _{z},$ the
Rashba and Dresselhaus couplings are interchanged\cite{Shen04prb}, 
\begin{eqnarray}
\alpha (\Pi _{x}\sigma _{x}-\Pi _{y}\sigma _{y}) &\rightarrow &\alpha (\Pi
_{x}\sigma _{y}-\Pi _{y}\sigma _{x}); \\
\beta (\Pi _{x}\sigma _{y}-\Pi _{y}\sigma _{x}) &\rightarrow &\beta (\Pi
_{x}\sigma _{x}-\Pi _{y}\sigma _{y}); \\
g_{s} &\rightarrow &-g_{s}.
\end{eqnarray}%
Therefore a system with Rashba coupling $\beta $, Dresselhaus coupling $%
\alpha $, and Lande g-factor $g_{s}$ is mapped on to a system with Rashba
coupling $\beta $, Dresselhaus coupling $\alpha $, and Lande g-factor $%
-g_{s} $. In particular, a system with only Dresselhaus coupling can be
mapped on to a system with only Rashba coupling and an opposite sign in $%
g_{s}$. At the symmetric point $\alpha =\beta $, $V_{so}^{2D}$ is invariant
under the transformation. $\alpha =-\beta $ is another symmetric point under
the transformation, $\sigma _{x}\rightarrow -\sigma _{y},$ $\sigma
_{y}\rightarrow -\sigma _{x},$ $\sigma _{z}\rightarrow -\sigma _{z}.$

\textit{Signs of the couplings. }Under the transformation, $\sigma
_{x}\rightarrow -\sigma _{x},$ $\sigma _{y}\rightarrow -\sigma _{y},$ $%
\sigma _{z}\rightarrow \sigma _{z},$we have $\alpha \rightarrow -\alpha $
and $\beta \rightarrow -\beta $. The eigenenergy spectrum is invariant under
the simultaneous sign changes of the two couplings. The eigenenergy spectrum
is even in $\alpha $ if $\beta =0$ and is even in $\beta $ if $\alpha =0$.

\textit{Charge conjugation.} Under the charge conjugation transformation, $%
-e\rightarrow e$, the magnetic moment of the carrier also changes its sign,
or effectively $g_{s}\rightarrow -g_{s}$ in Eq. (4). This transformation is
equivalent to the flip of the external magnetic field $B\rightarrow -B$.
Therefore, a system of hole carriers has the same physical properties as the
corresponding electron system except for possible directional changes in the
observables.

\subsection{RSHE in 2DEG with Rashba and Dresselhous couplings}

The Landau levels of 2DEG with a pure Rashba coupling in the absence of an
electric field was obtained over fourty years ago by Rashba, and studied by
a lot of authors.\cite{Rashba60} The Landau levels of 2DEG with a pure
Dresselhaus coupling can be obtained by symmetry as discussed in the last
subsection. However, the problem becomes more complicated in a system with
both the Rashba and Dresselhous couplings. The truncation and the
perturbation approximations have been applied to study the system and the
response to an electric field\cite{Shen05prb,Bao05prb}. This problem has
recently been solved by D. Zhang\cite{Zhang06jpa}. By introducing two boson
operators and using a unitary transformation, an exact solution can be
expressed as an infinite series in terms of the free Landau levels and other
parameters. Suing the result obtained from the exact solution it was
confirmed that a crossing between the energy eigenstates occur for larger
Rashba coupling, and the degeneracy may lead to the resonant spin Hall
conductance, which is consistent with the results obtained by using the
perturbation theory and truncation approximation.\cite%
{Shen04prl,Shen05prb,Bao05prb}

\subsection{RSHE in 2D hole gas}

The electronic structure of hole-doped systems near the $\Gamma $ point is
well described by the Luttinger Hamiltonian 
\begin{equation}
H_{L}=\frac{1}{2m}(\gamma _{1}+\frac{5}{2}\gamma _{2})\mathbf{p}^{2}-\frac{%
\gamma _{2}}{m}(\mathbf{p}\cdot \mathbf{J})^{2}
\end{equation}%
where $J$ is the total angular momemtun operator of $j=3/2$. If the system
breaks also the bulk inversion symmetry, a Rashba term $\beta (\mathbf{J}%
\times p)\cdot \hat{z}$ may be included in $H_{L}$. The Rashab term
introduces an energy splitting of the heavy holes in the 2-dimesnions, and
is reduced to the form of the so-called cubic Rashba term, $\gamma
(k_{-}^{3}\sigma _{+}+k_{+}^{3}\sigma _{-})$. In the presence of a
perpendicular magentic field, both the Luttinger model and cubic Rashba
model can be solved analytically. Recently, Zarea and Ulloa\cite{Zarea06prb}%
, and Ma and Liu\cite{Ma06prb} have obtained exact solutions in two
dimension, and used the Kubo-Greenwood formula to calculate the spin Hall
conductance. Neglecting the Zeeman splitting, they found that the spin Hall
conductance approaches to $9e/8\pi $ in the limit of a weak magnetic field
and low density of carriers, consistent with the result at the zero field.
This is in contrast to the linear Rashba coupling model of $H_{R}$, where
the spin Hall conductance vanishes in the absence of the Zeeman coupling. 
\cite{Rashba04prb} When the Zeeman splitting is taken into account in the
cubic Rashba system, it was found that a resonant spin Hall effect may occur
when the Fermi surface acrosses through the crossing point of the spectra as
the magnetic field decreases.\cite{Ma06prb}

\section{Edge spin current and resonant spin Hall effect}

It was well known that the edge state provides an alternative approach to
understand the quantum Hall effect in 2DEG. Here we discuss the edge spin
current and the resonant spin Hall effect in the Rashba system described by $%
H_{R}$ with a periodic boundary condition in the $x$ direction and two edges
along the $y-$ direction. The velocity operator in the Hilbert subspace of $%
y_{0}=p_{x}l_{b}^{2}/\hbar $ can be written as $v_{x}=(l_{b}^{2}/\hbar
)(\partial H/\partial y_{0})$ and thus the spin current operator 
\begin{equation}
j_{s}^{z}=l_{b}^{2}(\partial H/\partial y_{0}\sigma _{z}+\sigma _{z}\partial
H/\partial y_{0})/2.
\end{equation}%
Following the works by Laughlin,\cite{Laughlin81prb} Halperin \cite%
{Halperin82prb} and MacDonald and Streda \cite{MacDonald84prb} the\ total
spin Hall current in the filled Landau level ($n,s$) can be expressed as\cite%
{Bao05prb} 
\begin{eqnarray}
&&\left( j_{s}^{z}\right) _{n,s}=\int_{-L/2}^{+L/2}\frac{dy_{0}}{4\pi }\frac{%
\partial E_{n,s}(y_{0})}{\partial y_{0}}\left\langle \tau \right\vert \sigma
_{z}\left\vert \tau \right\rangle -  \notag \\
&&\sum_{n^{\prime },s^{\prime }}\int_{-L/2}^{+L/2}\frac{dy_{0}}{8\pi }%
(E_{n,s}(y_{0})-E_{n^{\prime },s^{\prime }}(y_{0}))\times  \notag \\
&&\left( \left\langle \tau \right\vert \partial _{y_{0}}\left\vert \tau
^{\prime }\right\rangle \left\langle \tau ^{\prime }\right\vert \sigma
_{z}\left\vert \tau \right\rangle -\left\langle \tau \right\vert \sigma
_{z}\left\vert \tau ^{\prime }\right\rangle \left\langle \tau ^{\prime
}\right\vert \partial _{y_{0}}\left\vert \tau \right\rangle \right)
\end{eqnarray}%
where $\tau ^{\prime }=\left( n^{\prime },y_{0},s^{\prime }\right) $ and $%
E_{n,s}(y_{0})$ are the energy spectrum. Without the spin-orbit coupling the
energy eigenstate satisfies $\left\langle \tau \right\vert \sigma
_{z}\left\vert \tau ^{\prime }\right\rangle =s\delta _{n,n^{\prime }}\delta
_{s,s^{\prime }}$ so that 
\begin{equation}
\left( j_{s}^{z}\right) _{n,s}=-seV/4\pi ,
\end{equation}%
which is only determined by the difference of charge voltage at the two
edges, 
\begin{equation}
E_{n,s}(L/2)-E_{n,s}(-L/2)=-eV.
\end{equation}%
Similar to the case for the charge Hall current\cite%
{Laughlin81prb,Halperin82prb}, the inclusion of impurities and Coulomb
interactions in the Hamiltonian does not affect the above result. The spin
Hall conductance displays a series of plateaus in the quantum Hall regime, 
\begin{equation}
G_{s}=(1-(-1)^{n})e/8\pi
\end{equation}%
corresponding to the quantum Hall conductance, $G_{c}=ne^{2}/h.$ In the
presence of spin-orbit coupling the states with different spins will be
mixed together, and the spin gradually deviates from the $z$- to $y$%
-direction and the spin Hall conductance varies with the effective Rashba
coupling or magnetic field through tuning the energy gap between the two
states especially near the Fermi level. The total spin Hall conductance can
be calculated numerically as a function of $1/B$ for a fixed chemical
potential \textit{\ assuming that the voltage drops only near the edges
hence the bulk state does not contribute to the total spin Hall current}.
The charge Hall conductance is found to be quantized as expected, and the
spin Hall conductance is of the order of $e/4\pi $ or $10^{-3}$ -- $%
10^{-4}e/4\pi $ depending on if odd or even number of Landau levels are
occupied. The spin Hall conductance is a function of the effective
spin-orbit coupling, which varies with the magnetic field. The resonant peak
appears only when the two degenerate bulk Landau levels crosses over a
special value of chemical potential with decreasing the magnetic field. The
spin Hall conductance for a fixed density of charge carriers can be
abstratced from the results for a fixed chemical potential if $L\gg l_{b}$.
The values of the spin Hall conductance are consistent with the bulk theory
for the fully filled Landau levels.\cite{Bao05prb}

\section{Discussions}

In this section we discuss an exact relation between the spin Hall current
and spin polarization and the effect of the disorder to the resonant spin
Hall effect. We briefly discuss the future experimental serach of the
resonant spin Hall effect.

\subsection{Spin Hall current and spin polarization}

For a Rashba system described by $H_{R}$ at $B=0$, it is now agreed that the
spin Hall conductance vanishes when the disorder effect is taken into account%
\cite{Sinova06ssc}. This may be understood by calculating the commutator 
\begin{eqnarray}
d\sigma _{x}/dt &=&[H_{R},\sigma _{x}]/i\hbar  \notag \\
&=&-\left( 4m\beta ^{2}/\hbar \right) J_{x}^{z}-g_{s}\mu _{B}B\sigma
_{y}/\hbar .
\end{eqnarray}%
Note that the above relation remains valid in the presence of non-magnetic
disorder. In a steady state the expectation value of $d\sigma _{x}/dt$
should vanish, and we see immediately from the above relation that the spin
Hall current vanishes in 2DEG described by a linear Rashba coupling in the
absence of magnetic field B.\cite{Rashba04prb} This can also be understood
from the point of view of spin force as examined by Shen\cite{Shen05prl} and
by Jin et al. \cite{Jin06jpa}. The spin-orbit coupling and the Zeeman
exchange coupling induce a spin force, which contains two parts: the
transverse force on a moving electron spin, $\mathbf{f}=4m^{2}\beta ^{2}%
\mathbf{J}^{z}\times \hat{z}$, \ with $\mathbf{J}^{z}$ the spin current
tensor component carried by the electron, and a spin force within the plane
induced by the Zeeman exchange coupling. The latter is relevant to the spin
polarization, $\mathbf{g}=-2\beta g_{s}m\mu _{B}B\left[ \sigma _{x}\hat{x}%
+\sigma _{y}\hat{y}\right] .$ If the disorder potential $V_{disorder}$ is
taken into account, in a steady state, the spin force must reach at balance, 
\begin{equation}
\frac{1}{i\hbar }\left\langle \left[ \frac{e}{c}\mathcal{A},H+V_{disorder}%
\right] \right\rangle =\left\langle \mathbf{f}+\mathbf{g}\right\rangle =0,
\end{equation}%
where $\mathcal{A}$ is the spin gauge vector potential caused by the
spin-orbit coupling. This result is independent of the non-magnetic disorder
and interaction because the spin gauge field commutes with non-magnetic
potential $V_{disorder}$. In this way we have established a relation between
spin current and spin polarization, 
\begin{eqnarray}
\ \left\langle J_{x}^{z}\right\rangle &=&-\frac{g_{s}\mu _{B}}{2m\alpha }%
B\left\langle \sigma _{y}\right\rangle  \notag \\
\left\langle J_{y}^{z}\right\rangle &=&+\frac{g_{s}\hbar \mu _{B}}{2m^{\ast
}\lambda }B\left\langle \sigma _{x}\right\rangle
\end{eqnarray}%
It becomes clear that the spin Hall current vanishes in the case of $B=0$ or 
$g=0$, which is consistent with previous results based on the vertex
correction calculations or non-equilibrium Green's function calculations.%
\cite{Inoue04prb} Note that the extrinsic or disorder contributions are
implicitly included in discussion of the spin force since it is required to
reach the balance or equilibrium for the system.

\subsection{Disorder effect and spin polarization}

The effects of disorder in 2DEG with Rashba coupling in a strong magnetic
field is not well understood at present. Nevertheless, it seems reasonable
to assume that the spin-orbit coupling does not change the effects of
disorder qualitatively. The strong magnetic field ensures extended states in
the Landau levels when the disorder is not sufficiently strong as evidenced
by the experimentally observed quantization of the charge Hall conductance.
One may assume that the disorder gives rise to broadening of the Landau
level so that the extended states in a Landau levels are separated in energy
from those in the next one by localized states. Inspection of the spin-orbit
coupling shows that Laughlin's gauge argument still holds\cite%
{Laughlin81prb,Halperin82prb}, and each Landau level with its extended
states completely filled contributes an amount of $e^{2}/h$ to the charge
Hall conductance. Thus one may conclude that the quantum Hall conductance
remains intact with the spin-orbit interaction, except at the special
degeneracy point.\cite{Shen04prl}

The disorder effect may be studied by numerical methods. Recently Bao and
Shen\cite{Bao06xxx} used the exact solutions of 2DEG in the absence of the
spin-orbit coupling and the disorder as the base function to numerically
solve the single particle problem by a truncation approximation. They
considered a short-range disorder potential $U(x,y)=\sum_{i}u_{i}\delta
(x-x_{i})\delta (y-y_{i})$ where $u_{i}\in (-u/2,u/2)$ and $(x_{i},y_{i})$
is distributed randomly in finite size of the system. They found that the
impurity scattering tends to suppress the spin Hall conductance. However, if
the disorder is not so strong the resonance remains intact.

\subsection{Experiment to observe resonant spin Hall effect}

As we discussed in this brief review, the theoretical studies predict a
resonant spin Hall effect in two-dimensional electron or hole systems with
spin-orbit coupling in a strong magnetic field and at low temperatures. It
is likely due to the difficulty in detecting the spin current in these
extreme conditions that no experiments have been carried out to observe the
resonant effect. It is our view that the resonance should be measurable. The
most promising way to observe the effect is to detect the spin polarization
or the spin susceptibility. As we discussed in Section IV B, the spin
polarization is proportional to the spin Hall current. The resonant effect
may be best observed in the rapid change of the direction of spin
polarization as magnetic field sweeps through the resonant field in the
Rashba sytems. Since the change of the spin polarization in the system only
occurs at the Landau levels near the Fermi energy, the effect will be most
pronounced if the resonance occurs at the lower Landau levels.

\begin{acknowledgments}
We would like to thank M. Ma, X. C. Xie, Y. J. Bao, and Y. Q. Li for many
discussions on the subject. This work was supported in part by the Research
Grant Council in Hong Kong with Grant No. HKU 7039/05P and HKU 7042/06P.
\end{acknowledgments}

\end{document}